\documentclass[twocolumn]{autart}
\input epsf
\usepackage{cite,epsfig}
\usepackage{graphicx,amsmath,amssymb,remark,color}
\usepackage[T1]{fontenc}
\usepackage[latin1]{inputenc}
\usepackage{graphicx}
\usepackage{amssymb}
\usepackage{amsmath}
\usepackage{dcolumn}
\usepackage{bm}
\usepackage{color}
\usepackage{multicol}
\usepackage{algorithm}
\usepackage{multirow}
\usepackage{subfigure} 
\input epsf
\usepackage{cite,epsfig}
\usepackage{graphicx,amsmath,amssymb,remark,color}
\newcommand{\EQ}{\begin{eqnarray}}
\newcommand{\EN}{\end{eqnarray}}
\newcommand{\EQQ}{\begin{eqnarray*}}
\newcommand{\ENN}{\end{eqnarray*}}
\def\widebar{\accentset{{\cc@style\underline{\mskip10mu}}}}

 \makeatletter
\newcommand*\bigcdot{\mathpalette\bigcdot@{.5}}
\newcommand*\bigcdot@[2]{\mathbin{\vcenter{\hbox{\scalebox{#2}{$\m@th#1\bullet$}}}}}
\makeatother

\DeclareMathAlphabet{\matheur}{U}{eur}{m}{n}
\DeclareMathAlphabet{\matheurb}{U}{eur}{b}{n}
\DeclareMathAlphabet{\matheus}{U}{eus}{m}{n}
\DeclareMathAlphabet{\matheuf}{U}{euf}{m}{n}
\renewcommand{\t}{^{\mbox{\tiny\sf T}}}

\newcommand{\cal}{\mathcal}
\newcommand{\eproof}{\hfill\rule{2mm}{2mm}}

\newcommand{\bremark}{\begin{remark}
\begin{rm}}
\newcommand{\eremark}{ \end{rm}
\end{remark} }
\newcommand{\btheorem}{\begin{theorem} \begin{it}}
\newcommand{\etheorem}{\end{it} \hfill \rule{1mm}{2mm}
\end{theorem} }
\newcommand{\blemma}{\begin{lemma} \begin{it} }
\newcommand{\elemma}{ \end{it} \hfill\rule{1mm}{2mm}
\end{lemma} }
\newcommand{\bcorollary}{\begin{corollary} \begin{it} }
\newcommand{\ecorollary}{ \end{it} \hfill\rule{1mm}{2mm}
\end{corollary} }
\newcommand{\bdefinition}{\begin{definition} }
\newcommand{\edefinition}{ \hfill\rule{1mm}{2mm}
\end{definition} }
\newcommand{\bproposition}{\begin{proposition} }
\newcommand{\eproposition}{\hfill \rule{1mm}{2mm}
\end{proposition} }
\newcommand{\bexample}{\begin{example} \begin{rm}}
\newcommand{\eexample}{ \end{rm} \hfill\rule{1mm}{2mm}
\end{example} }

\newcommand{\basm}{\begin{assumption} \begin{rm} }
\newcommand{\easm}{ \end{rm} \hfill\rule{1mm}{2mm}
\end{assumption} }

\begin{document}

\newtheorem{theorem}{\sf\bfseries Theorem}[section]
\newtheorem{lemma}{\sf\bfseries Lemma}[section]
\newtheorem{coro}{\sf\bfseries Corollary}[section]
\newtheorem{definition}{\sf\bfseries Definition}[section]
\newtheorem{remark}{\sf\bfseries Remark}[section]
\newtheorem{corollary}{\sf\bfseries Corollary}[section]
\newtheorem{proposition}{\sf\bfseries Proposition}[section]
\newtheorem{example}{\sf\bfseries Example}[section]
\newtheorem{assumption}{\sf\bfseries Assumption}

\begin{frontmatter}

\title{Cooperative Label-Free Moving Target Fencing for Second-Order Multi-Agent Systems with Rigid Formation}

\author[hust]{Bin-Bin Hu},
\author[hust]{Hai-Tao Zhang$^*$}, 
\author[vict]{Yang Shi}

\address[hust]
{School of Artificial Intelligence and Automation \\
Key Laboratory of Image Processing and Intelligent Control \\
State Key Lab of Digital Manufacturing Equipment and Technology\\
Huazhong University of Science and Technology\\
Wuhan 430074, P.R.~China \\
Email:  {\tt binbinhu1995@gmail.com, zht@mail.hust.edu.cn}}  

\address[vict]{Department of Mechanical Engineering\\
University of Victoria\\
Victoria, BC V8W 2Y2, Canada \\
Email:  {\tt yshi@uvic.ca}\\
$^*$ Corresponding author
}

\thanks{This work was supported by in part by the National Natural Science
Foundation of China under Grants 62225306, U2141235, 51729501, in part by the National Natural Science Foundation of Hubei Province under Grant 2019CFA005.} 
 

\begin{abstract}
This paper proposes a label-free controller for a second-order multi-agent system to cooperatively fence a moving target of variational velocity into a convex hull formed by the agents whereas maintaining a rigid formation. Therein, no label is predetermined for a specified agent. To attain a rigid formation with guaranteed collision avoidance, each controller consists of two terms: a dynamic regulator with an internal model to drive agents towards the moving target merely by position information feedback, and a repulsive force between each pair of adjacent agents. Significantly,  sufficient conditions are derived to guarantee the asymptotic stability of the closed-loop systems governed by the proposed fencing controller.  Rigorous analysis is provided to eliminate the strong nonlinear couplings induced by the label-free property. Finally, the effectiveness of the controller is substantiated by numerical simulations.
\end{abstract}

\end{frontmatter}

\section{Introduction}

Cooperative control of multi-agent systems (MASs) is motivated by high coordination of the bird flocks, fish schools, insect colonies and mammal herds in the natural world, which can be applicable to  a large volume of industry, engineering and social networked systems by virtues of high efficiency, large coverage and low cost.  Emergence of high cooperation is a fascinating topic in the MAS control research area all along. Tremendous progress has been witnessed in recent years, including formation regulation \cite{olfati2006flocking,zhang2010general},  synchronization/consensus  protocols~\cite{chen2013remark,jin2020event} containment control \cite{meng2010distributed,yang2021passive}, and circular-motion control~\cite{chen2011no,sun2018circular}.

Along the research line of collective circular motion control, a target enclosing issue has recently attracted increasing attention. Most of the pioneer efforts devoted to the MAS target enclosing problem have focused on containing a target within the moving trajectories of well-informed agents \cite{lan2010distributed,kim2007cooperative,zheng2015enclosing}, who have access to the target's information. However, such a situation is not often encountered in real applications, and hence a more practical cooperative strategy was proposed in \cite{yu2016distributed} to encircle a target known to only a partial of agents.

Even though a target could be encircled by moving trajectories in \cite{lan2010distributed,kim2007cooperative,zheng2015enclosing,yu2016distributed}, the consistent enclosing of a target at every moment cannot be always guaranteed. That motivates another interesting research line, namely, target-fencing problem, which means that the target is fenced (i.e., surrounded / enclosed) by the convex hull of all the agents all along. Among the initial works of MAS fencing control, all the agents are label-fixed, i.e., their labels, neighbors and relative distribution are predetermined. 
Such a target-fencing control problem has been tackled for first-order MASs \cite{chen2010surrounding} and second-order MASs \cite{shi2015cooperative}. Afterwards, it was extended to multi-targets fencing scenario for second-order nonlinear MASs in \cite{Hu2020Multiple} and even multiple unmanned surface vessels \cite{Hu2021Distributed}.

Nevertheless, such kind of label-fixed fencing controllers cannot fulfill increasingly complicated missions in unpredictable environment. 
For instance, with the increasingly complicated missions, label-fixed controllers need to recalculate the predetermined inter-agent relative distribution every time, which will increase calculation burden and thereby decrease the cooperation efficiency. As for unpredictable environments, label-fixed controllers may take a longer time and consume more energy to attain cooperation in some specific situations. For instance, if the initial position of a specific agent~$1$ is far behind all the other ones whereas its predetermined relative position is in front of all the others, then agent~$1$ needs to trudge across all the others to the prescribed position. Last but not least, if some agents break down suddenly during the fencing process, label-fixed controllers  
can not form a fencing formation any longer since all the inter-agent relative position are predetermined in advance.
This motivates the development of a more practical label-free fencing controller, where no label is predetermined for a specified agent \cite{sakurama2020multi}. To this end, a limit-cycle-based protocol was proposed in \cite{wang2017limit} to fence a stationary target. A hierarchical structure with an output regulation method was established in \cite{Liubin2018surounding} to surround a stationary target. A cooperative controller consisting of attractive, repulsive, and rotational inter-agent forces was developed in~\cite{chen2019cooperative} to fence a specified stationary target. Afterwards, an interesting 
problem to fence a target with a constant velocity was addressed for first-order MASs \cite{kou2021cooperative1}, second-order MASs \cite{kou2021cooperative,hu2021distributed2} and multiple unmanned surface vessels \cite{hu2021bearing}.

So far, most of existing works \cite{wang2017limit,Liubin2018surounding,chen2019cooperative} only considered label-free fencing for first-order MASs and have not systematically considered the formation evolution during entire fencing processes, which is however essential in practice, such as unmanned-system convey protection, reconnaissance, patrol, etc. Although a few recent works  \cite{kou2021cooperative1,kou2021cooperative,hu2021distributed2,hu2021bearing} studied the rigid formation with a constant-velocity target, a more challenging scenario of fencing a moving target with variational velocity still remains a dilemma. Thus, it becomes an urgent yet challenging mission to propose a label-free controller for second-order MASs to achieve collision-free rigid-formation fencing for a variational-velocity target. Hereby, the main contribution of this paper is summarized.

\begin{enumerate}
\item Develop a label-free controller for a second-order MAS to cooperatively  fence a motional target of variational velocity within their convex hull whereas maintaining a rigid formation.

\item  Guarantee inter-agent collision avoidance and convergence of a nonlinear MAS subject to complicated dynamics, strong couplings and time-varying network topologies, simultaneously.

\end{enumerate}

Technically speaking, the main difficulty of this paper is the strong nonlinear couplings induced by the label-free property with a variational-velocity target. The novelty of this paper is two-fold. First of all, unlike relevant prior 
label-free fencing works for a static target \cite{wang2017limit,Liubin2018surounding,chen2019cooperative} and a constant-velocity target \cite{kou2021cooperative1,kou2021cooperative,hu2021distributed2,hu2021bearing}, the present study regards the target as an exosystem. Thereby, it proposes a label-free controller consisting of a dynamic regulator and an inter-agent repulsion to address a more challenging fencing problem with a variational-velocity target.
Secondly, by inserting an inter-agent repulsion into the internal model, this paper achieves a  
rigid formation with guaranteed collision avoidance. Still worth-mentioning is that, by utilizing a dynamic regulator with the internal model, the present method can cope with the previous constant-velocity target fencing problem as a specific case.

The remainder of this paper is organized as follows.
Section~II provides the preliminaries and the main problem addressed by the paper. Section~III derives the target-fencing control law and then presents the main technical results. 
Numerical simulations are conducted in Section~IV to  substantiate the effectiveness of the presented scheme. 
Finally, conclusion is drawn in Section~V.

Throughout the paper,  $\mathbb{R}$ and $\mathbb{R}^{+}$ denote the real number and positive real number spaces, respectively. $\mathbb{R}^{n}$ denotes $n$-dimensional Euclidean space, $\|v\|$ is the Euclidean norm of a vector $v$. $\otimes$ represents the Kronecker product.

\section{Problem Formulation}
Consider an MAS consisting of $n$ agents, of which each agent is governed by second-order dynamics in the Cartesian coordinates, 
\begin{align}
\label{kinetic_F}
\dot{x}_i(t) &=v_i(t)\nonumber\\
\dot{v}_i(t) &=u_i(t),
\end{align}
where $x_i(t), v_i(t)\in \mathbb{R}^2$ denote the position and velocity of agent $i$, respectively, and $u_i(t)\in \mathbb{R}^2$ the control input. The topology of the MAS is represented by~$\cal G=(\cal V, \cal E)$, where ${\cal V}= \{1, 2, \dots, n\}$ is the node set and $\cal E\subseteq \cal V \times \cal V$ the edge set. 
$\mathcal {N}_i$ is defined as the sensing neighborhood set of agent~$i$ in $\cal V$, i.e.,
\begin{align}
\label{sensing_neighbors}
\mathcal N_i(t):=\{k\in {\cal V},k\neq i \; | \: \|x_{i,k}(t)\| \leq R\}
\end{align} 
with a detectable range $R\in(r, \infty)$ and a specified safe distance $r\in\mathbb{R}^+$. $x_{i,k}(t):=x_i(t)-x_k(t), i\neq k\in \cal V$ are the relative position between agents $i$ and $k$. 
Due to the limited and changeable relative locations $\|x_{i,k}\|$, the neighborhood set $\mathcal N_i$ of agent $i$ is time-varying, which explicitly shows that the proposed topology $\cal G$
may keep changing as well. It is more challenging than the scenarios of fixed topologies with predetermined agent labels (see, e.g., \cite{chen2010surrounding,shi2015cooperative,Hu2021Distributed,Hu2020Multiple}).

Consider a moving target satisfying 
\begin{align}
\label{kinetic_L}
\begin{bmatrix}
 \dot{x}_d (t) \\
  \dot{v}_d (t)
\end{bmatrix}=S
\begin{bmatrix}
x_d(t)\\
v_d(t)
\end{bmatrix}
\end{align}
\begin{figure}[!htb]
  \centering
  \includegraphics[width=\hsize]{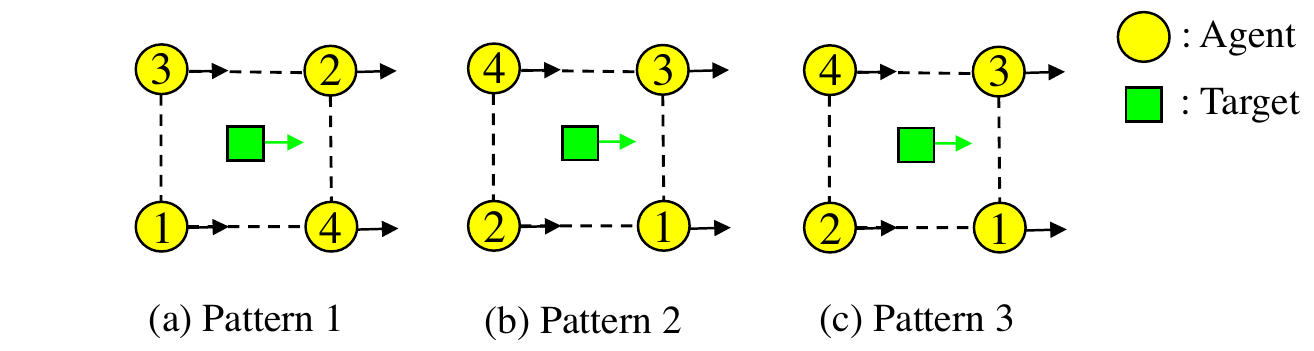}
  \caption{Example of label-free fencing for a motional target.} 
  \label{definittion}
\end{figure}
with the dynamic matrix 
\begin{align}
\label{matrix_target}
S=\begin{bmatrix}
0 &1 \\
s_1 &0
\end{bmatrix}\otimes I_2.
\end{align}
Here, $s_1\leq0$ is a constant, $x_d\in \mathbb{R}^2$, $v_d\in \mathbb{R}^2$ are the position and velocity of the target, respectively.
If $s_1=0$ in \eqref{matrix_target}, the target in \eqref{kinetic_L} moves with a constant velocity, see e.g.,  \cite{kou2021cooperative}. If $s_1<0$, the target moves 
periodically with a variational velocity.

Define ${\rm co}(x)$ as the convex hull of all the agents $x=[x_1,\cdots,x_n]\t$, i.e.,   
\begin{equation}
\label{convexhull}
        {\rm co}(x) := \left\{\sum_{i \in \cal V}\lambda_i x_i: \lambda_i \geq 0,\forall i \enspace \text{and}  \enspace \sum_{i \in\cal V} \lambda_i =1\right\},
\end{equation}
one has that the distance from target to the convex hull ${\rm co}(x)$ is calculated as $P_{x_d}(x) :=  \min_{s \in    {\rm co}(x)} \|x_d - s\|$ (see, e.g. \cite{chen2019cooperative}).

Next, we can give the following definition.

\begin{definition}
\label{rigid}
(Label-free rigid-formation fencing)~\cite{chen2019cooperative}: An MAS~$\mathcal V$ with dynamic~\eqref{kinetic_F} asymptotically fences a target with dynamic~\eqref{kinetic_L} into a convex hull \eqref{convexhull} formed by the arbitrary labelled agents whereas maintaining a collision-free rigid formation, if the following three claims are fulfilled,
\begin{align}
\label{defnition_rigid1}
&1)~\lim_{t\rightarrow\infty}P_{x_d(t)}(x(t)) = 0,\nonumber\\
&2)~\lim_{t\rightarrow\infty}v_i(t)-v_d(t)=0, \forall i\in \cal V, \nonumber\\
&3)~\|x_{i,k}(t)\|>r, \forall t\geq 0, \; \forall i\neq k \in \cal V.
\end{align}
\end{definition}

In Definition~\ref{rigid}, Claim 1) indicates that the target $x_d$ is fenced into a convex hull ${\rm co}(x)$ by arbitrary labelled agents $x_i, i\in\mathcal V$. From $\lim_{t\rightarrow\infty}v_i(t)-v_d(t)= 0$ in Claim~2), it can be deduced that $\lim_{t\rightarrow\infty}x_i(t)-x_d(t)= d_i$ with a constant vector $d_i\in\mathbb{R}^2$, which implies that the maintenance of relative positions among the agents and target, and hence the pattern of the MAS is guaranteed fixed. Claim 3) assures that the distance $\|x_{i,k}(t)\|$ between any pair of agents is always larger than $r$, i.e., inter-agent collision avoidance is guaranteed.    
\begin{figure}[!htb]
  \centering
  \includegraphics[width=\hsize]{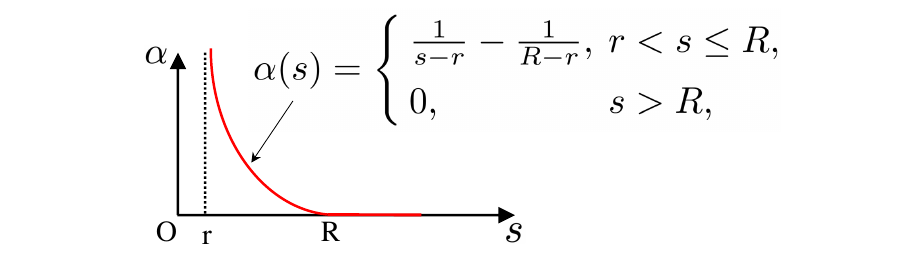}
  \caption{Illustration of the potential function $\alpha(s)$ in \eqref{form_potentional_function}. } 
  \label{Illustration_function}
\end{figure}
From Claims 1) and 3), it only achieves the fencing. Claim 2) is an extra requirement to realize rigid-formation fencing. Moreover, it is observed that Claims 1)-3) do not predetermine fixed labels and inter-agent relative positions for specific agents explicitly, which implies the labels of agents in the fencing patten are flexible, i.e., label-free fencing.

To show the main problem addressed by the paper more vividly, we illustrate an example of label-free fencing in Fig.~\ref{definittion}. 
Therein, all of the patterns~1-3 satisfy the fencing of a motional target, which also implies that no label is required for specified agents.

Now, it is ready to introduce the main problem addressed by the paper as below.

\textbf{Problem 1.}
Design a cooperative label-free controller 
\begin{align}
\label{virtual_law}
u_{i}:=f(x_i, v_i, x_d, x_k), i \in\mathcal V, k\in\mathcal N_i, 
\end{align}
for an MAS~\eqref{kinetic_F} with a variational-velocity target \eqref{kinetic_L} to achieve label-free rigid-formation fencing in Definition~\ref{rigid}.

\section{Main Results}
Let the relative position between an arbitrary agent $i\in \cal V$ and the target be 
\begin{align}
\label{e_error}
e_i(t):=x_i(t)-x_d(t). 
\end{align}

Now, design a cooperative controller with a dynamic feedback regulation for $i\in\cal V$ as below, with $(t)$ omitted for conciseness,
 \begin{align}
\label{control_i1}
u_i= &-\Big(\begin{bmatrix}
        k_1 & k_2
        \end{bmatrix}\otimes I_2\Big)
        \begin{bmatrix}
        x_i\\
        v_i\\
        \end{bmatrix}-
        \Big(
        \begin{bmatrix}
        k_3 & k_4
        \end{bmatrix}\otimes I_2\Big)
        \begin{bmatrix}
        \epsilon_i\\
        \zeta_i\\
        \end{bmatrix} \nonumber \\
        &+k_5\sum_{k\in\mathcal N_i} \alpha(\|x_{i,k}\|)\frac{x_{i,k}}{\|x_{i,k}\|}, \nonumber \\ 
 \begin{bmatrix}
        \dot{\epsilon}_i\\
        \dot{\zeta}_i\\
 \end{bmatrix}= &\bigg( \begin{bmatrix}
                             0 & 1\\
                             s_1 & 0\\                             
                             \end{bmatrix}\otimes I_2\bigg)
                              \begin{bmatrix}
       		             \epsilon_i\\
                              \zeta_i\\
                              \end{bmatrix}+
                              \bigg(
                              \begin{bmatrix}
                             0 \\
                             1 \\                             
                             \end{bmatrix}\otimes I_2\bigg)\nonumber\\
                             &\bigg(e_i-k_5\sum_{k\in\mathcal N_i} \alpha(\|x_{i,k}\|)\frac{x_{i,k}}{\|x_{i,k}\|}\bigg),         
\end{align}
where $k_{1},k_{2}, k_{3}, k_{4}, k_{5}\in\mathbb{R}^+$ are all positive parameters, and $\epsilon_i, \zeta_i\in\mathbb{R}^2$ the states of the internal model in \eqref{control_i1} for agent~$i$, respectively.
$\mathcal N_i$ is given in \eqref{sensing_neighbors}. $\sum_{k\in\mathcal N_i} \{\alpha(\|x_{i,k}\|){x_{i,k}}/{\|x_{i,k}\|}\}$ denotes the inter-agent repulsion, where the continuous function $\alpha(\cdot)$ (see, e.g.~\cite{chen2019cooperative})
satisfying 
\begin{align}
\label{alpha}
\alpha(s)=0, \forall s\in[R, \infty), 
\lim_{s\rightarrow r^{+}}\alpha(s)=\infty
\end{align}
with the detectable range $R\in(r, \infty)$ in and the specified safe distance $r$ in \eqref{sensing_neighbors}. Throughout the paper, denote
\begin{align}
\label{neat_eta}
\eta_i:=\sum_{k\in\mathcal N_i} \{\alpha(\|x_{i,k}\|)\frac{x_{i,k}}{\|x_{i,k}\|}\}
\end{align}
for conciseness of presentation.

\begin{remark}
The $\alpha(s)$ in \eqref{alpha} is monotonically decreasing when $s\in(r, R]$, which equals $0$ when $s\in(R, \infty)$. It implies that $\alpha(s)$ is Lipschitz continuous in $(r, \infty)$. As shown in Fig.~\ref{Illustration_function}, an example of $\alpha(\cdot)$ satisfying~\eqref{alpha} is given as below,
\begin{equation}
\label{form_potentional_function}
\alpha(s)=
\left\{
\begin{array}{llr}
\frac{1}{s-r}-\frac{1}{R-r}, & r<s\leq R,\\
0, & s>R.
\end{array}
\right.
\end{equation}

\end{remark}

\begin{remark}
From the controller~\eqref{control_i1}, the position state $x_d$, albeit available to each agent, is not the final convergent steady state, which is different from centralized controllers (like star topology). In other words, the steady states of agents in~\eqref{control_i1} are distributedly calculated by the attraction domain of the target and the local interactions of agents,
which is well accepted in most label-free fencing papers \cite{wang2017limit,Liubin2018surounding,chen2019cooperative,kou2021cooperative1,kou2021cooperative}.
Moreover, when the target position $x_d$ is only available to a small partial of the agents in real applications, the position $x_d$ could be transferred to each agent via communication network in finite time,  see, e.g., \cite{sun2014finite}.
\end{remark}


\begin{remark}
The internal model $\epsilon_i, \zeta_i$ in Eq.~\eqref{control_i1} represents a dynamic compensator to compensate the variational velocity of the target $v_d(t)$ with position-only measurement, which thus guarantees the solution of the augmented system. Moreover, by inserting the inter-agent repulsion $k_5\sum_{k\in\mathcal N_i} \alpha(\|x_{i,k}\|){x_{i,k}}/{\|x_{i,k}\|}$ into the internal model states $\epsilon_i, \zeta_i$, a necessary ingredient is yielded to achieve a rigid formation with a variational-velocity target during the fencing process, which will be proved in Lemma~\ref{rigid_property}. The dynamics matrices $G_1, G_2$ in Eq.~\eqref{control_i1} can be calculated by minimal characteristic polynomial of the dynamic matrix $S$ in \eqref{matrix_target} according to Lemma~\ref{p_copy_model}.
\end{remark}

\begin{remark}
The labels in the label-free controller $u_i$~\eqref{control_i1} are just utilized to distinguish from other agents. Since the form of the controller~\eqref{control_i1} is identical for each agent, there is no difference if any pair of individual controllers exchange their labels. By contrast, however, in the label-fixed design, each controller predetermines specific and fixed relative positions among agents, which is thus different from any other controllers.
\end{remark}

Next, we will prove that the closed-loop MAS governed by \eqref{kinetic_F}, \eqref{kinetic_L}, \eqref{matrix_target} and \eqref{control_i1} satisfies the following property.

\begin{itemize}

\item[{\bf P1}:] An MAS $\mathcal V$ achieves Claims 1)-3) in collision-free rigid-formation fencing  (see Definition~\ref{rigid}).
 
\end{itemize}

 To this end, the following conditions are required.

 \begin{itemize}

\item[{\bf C1}:] The initial positions of the agents $\mathcal V$ satisfy
$\|x_{i,k}(0)\|>r,  \; \forall i\neq k \in \cal V$;

\item[{\bf C2}:] The parameters $k_1, k_2, k_3, k_4\in\mathbb{R}^+$ in \eqref{control_i1} satisfy  
\begin{align}
\label{condition_C2}
k_2=k_4\frac{k_1+s_1-1}{k_3},  ~{k_1-k_3+s_1-1}>0.
\end{align}

\item[{\bf C3}:] The dynamic matrix $S$ in \eqref{matrix_target} of the target is available to all the agents.
\end{itemize}

\begin{remark}
Condition C1 is necessary for collision avoidance design. Condition C2 simultaneously guarantees fencing property, collision avoidance and rigid formation.
Condition C3 is necessary to design the internal model in \eqref{control_i1}, whose reason is given below. 
The matrix~$S$ is assumed to
be known to each agent to calculate the dynamic matrices $\mathcal G_1, \mathcal G_2$ of the internal model in Lemma~\ref{p_copy_model} later.
Moreover, since the matrix $S$ of the target is available to the agents, the modal composition of the target velocity $v_d$ (e.g., the structure and frequency of the velocity variation: $f=1/\sqrt{-s_1}$Hz if $s_1<0$, and $f=0$Hz if $s_1=0$) is known to the agents as well. Then, according to the matrix $S$ and the available position of the target $x_d(t)$, the real-time variational velocity $v_d(t)$ can be calculated bleow.
It follows from the dynamics of the target in Eq.~\eqref{kinetic_L} that $x_d(t), v_d(t)$ evolve below 
\begin{align}
\label{partial_inequal3}
\begin{bmatrix}
x_d(t)\\
v_d(t)
\end{bmatrix}=&
\Phi(t)
\begin{bmatrix}
x_d(0)\\
v_d(0)
\end{bmatrix}
\end{align}
with the initial position and velocity $x_d(0), v_d(0)$, and  
\begin{align}
&\Phi(t)=
\begin{bmatrix}
\label{dynamic_matrix}
\cos(\sqrt{-s_1}t) & \frac{\sin(\sqrt{-s_1}t)}{\sqrt{-s_1}}\\
-\sqrt{-s_1}\sin(\sqrt{-s_1}t) & \cos(\sqrt{-s_1}t)
\end{bmatrix}\otimes I_2, \mbox{if}~s_1<0, \nonumber\\
&\Phi(t)=
\begin{bmatrix}
1 & t\\
0 & 1
\end{bmatrix}\otimes I_2, \mbox{if}~s_1=0.
\end{align}
Here, the structure of $\Phi(t)$ in Eq.~\eqref{dynamic_matrix} and its frequency ($f=1/\sqrt{-s_1}$Hz if $s_1<0$, and $f=0$Hz if $s_1=0$) denote the modal composition of the target velocity $v_d(t)$.
Then, it follows from Eqs.~\eqref{partial_inequal3} and \eqref{dynamic_matrix} that the initial velocity $v_d(0)$ can be calculated with $\Phi(t)$, $x_d(0)$, and $x_d(t)$, which implies that the real-time velocity $v_d(t)$ can be calculated by $\Phi(t), x_d(0), v_d(0)$ as well. 
By regarding the motional target as an exosystem, the cooperative target fencing in Problem~1 is transformed into a cooperative regulation problem,
where condition C3 is a well-accepted assumption in the design of cooperative regulation works \cite{su2011cooperative,su2012cooperative}. 
\end{remark}

Before presenting the main technical results, it is necessary to introduce some preliminaries.  

\begin{lemma}
\label{lemma_ORP}
\cite{huang2004nonlinear} Consider a linear time-invariant system governed by 
\begin{align*}
\dot{x}&=\mathcal Ax+\mathcal Bu+\mathcal E\nu,\nonumber\\
e&=\mathcal Cx+\mathcal Du+\mathcal F\nu, \nonumber\\
\dot{\nu}&=\mathcal S\nu,
\end{align*}
and a dynamic state feedback controller 
\begin{align*}
u=& \mathcal K_1x+\mathcal K_2z,~\dot{z}=\mathcal G_1z+\mathcal G_2 e,
\end{align*}
where $x\in\mathbb{R}^n, u\in\mathbb{R}^m, e\in\mathbb{R}^p, z\in \mathbb{R}^{n_z}, v\in\mathbb{R}^q$ are the system state, input state, regulated output, internal model state and exosystem signal, respectively, and $\mathcal A\in\mathbb{R}^{n\times n}, \mathcal  B\in\mathbb{R}^{n\times m}, \mathcal  C\in\mathbb{R}^{p\times n}, \mathcal D\in\mathbb{R}^{p\times m}, \mathcal E\in\mathbb{R}^{n\times q}, \mathcal F\in\mathbb{R}^{p\times q}, \mathcal S\in\mathbb{R}^{q\times q}, \mathcal K_1\in\mathbb{R}^{m\times n}, \mathcal K_2\in\mathbb{R}^{m\times n_z}, \mathcal G_1\in\mathbb{R}^{n_z\times n_z}, \mathcal G_2\in\mathbb{R}^{n_z\times p}$. Assume $\mathcal S$ has no eigenvalue with negative real parts. If $(\mathcal G_1, \mathcal G_2)$ incorporates a minimal p-copy internal model of the matrix $\mathcal S$ and the matrix 
\begin{align*}
\mathcal A_c=\begin{bmatrix}
\mathcal A+\mathcal B\mathcal K_1 & \mathcal B\mathcal K_2\\
\mathcal G_2\mathcal (C+\mathcal D\mathcal K_1) & \mathcal G_1+\mathcal G_2\mathcal D\mathcal K_2
\end{bmatrix}
\end{align*}
is Hurwitz, then the following Sylvester matrix equation has a unique solution $\mathcal X\in\mathbb{R}^{n\times q}, \mathcal Z\in\mathbb{R}^{n_z\times q}$ as 
\begin{align}
\label{lemma_matrix_equation}
\mathcal X_c\mathcal S=&\mathcal A_c
\mathcal X_c+\mathcal B_c,\nonumber\\
\mathbf{0}=&\mathcal C_c\mathcal X_c+\mathcal D_c
\end{align}
with 
\begin{align*}
\mathcal X_c=\begin{bmatrix}
\mathcal X\\
\mathcal Z
\end{bmatrix}, 
\mathcal B_c=\begin{bmatrix}
\mathcal E\\
\mathcal G_2 \mathcal F
\end{bmatrix},
\mathcal C_c=\begin{bmatrix}
\mathcal C & \mathcal D
\end{bmatrix},
\mathcal D_c=\begin{bmatrix}
\mathcal F
\end{bmatrix}.
\end{align*}
\end{lemma}

\begin{lemma} \cite{huang2004nonlinear}
\label{p_copy_model}
Consider an arbitrary square matrix $\mathcal S\in\mathbb{R}^{q\times q}$ and a regulated output $e\in\mathbb{R}^p$, a pair of matrices $(\mathcal G_1, \mathcal G_2)$ is said to incorporate a minimal p-copy internal model of the matrix~$\mathcal S$ if the matrices $\mathcal G_1, \mathcal G_2$ can be described below 
\begin{align*}
\mathcal G_1=&\mbox{block diag}~[\beta_{1}, \beta_{2}, \dots, \beta_{p}]\in\mathbb{R}^{pq\times pq},\nonumber\\
\mathcal G_2=&\mbox{block diag}~[\sigma_1, \sigma_2, \dots, \sigma_p]\in\mathbb{R}^{pq},
\end{align*} 
where $p$ refers to the dimension of the regulated output $e$, $``\mbox{block diag}"$ represents a block diagonal matrix,  $\beta_i\in\mathbb{R}^{q\times q}$ is a constant square matrix and $\sigma_i\in\mathbb{R}^{q}$ is a constant column vector such that the following two conditions are satisfied.

(1) ($\beta_i, \sigma_i$) are controllable;

(2) The minimal characteristic polynomial of $\mathcal S$ divides the characteristic polynomial of $\beta_i$.


Moreover, let the minimal characteristic polynomial of $\mathcal S$ be
$$S(\lambda):=\lambda^{n_m}+\alpha_1\lambda^{n_m-1}+\cdots+\alpha_{n_m-1}\lambda+\alpha_{n_m},$$
one has that $\beta_i, \sigma_i, i=1,2, \dots, p$ can be designed as follows,
\begin{align*}
\beta_i=&\begin{bmatrix}
0 & 1 & \cdots & 0\\
0 & 0 & \cdots & 1\\
\vdots & \vdots &\vdots & \vdots\\
0 & 0 & \cdots & 1\\
-\alpha_{n_m} & -\alpha_{n_m-1} & \cdots &-\alpha_1
\end{bmatrix},
\sigma_i=
\begin{bmatrix}
0\\
0\\
\dots\\
0\\
1
\end{bmatrix},
\end{align*}
which satisfy the aforementioned conditions~(1), (2) of the p-copy internal model.
\end{lemma}

\begin{remark}
As illustrated in Lemma~\ref{p_copy_model}, the matrix $\mathcal G_1$ contains all the eigenvalues of the matrix~$\mathcal S$, and the matrix $\mathcal G_2$ makes the pair matrices $(\mathcal G_1, \mathcal G_2)$
controllable, which thus can guarantee the solution of the Sylvester matrix equation in Eq.~\eqref{lemma_matrix_equation} of Lemma~\ref{lemma_ORP}.
\end{remark}

\begin{lemma}
\label{lemma_schur}
(Schur Complement \cite{boyd1994linear}).  The linear matrix inequality
\begin{align*}
 \begin{bmatrix}
T(p) & W(p)\\
W\t(p) & R(p)\\
\end{bmatrix}>0
\end{align*}
with $T(p)=T\t(p)$ and $R(p)=R\t(p)$, is equivalent to either of the following statement:
\begin{align*}
1.~T(p)>0,~R(p)-W\t(p)T^{-1}(p)W(p)>0,\nonumber\\
2.~R(p)>0,~T(p)-W(p)R^{-1}(p)W\t(p)>0.
\end{align*} 
\end{lemma}
For convenience of the readers, we divide the main technical results into three steps, i.e., the fencing, inter-agent collision avoidance and rigid formation.
First, a lemma concerning fencing property in Step 1 is provided.
\begin{lemma}
\label{fenceing_property}
Under condition C3, an MAS $\mathcal V$ governed by \eqref{kinetic_F}, \eqref{control_i1} is able to collectively fence a motional target governed by \eqref{kinetic_L}
i.e., Claim 1) in Eq.~\eqref{defnition_rigid1} 
if and only if (iff) the control gains $k_1, k_2, k_3, k_4  \in \mathbb R^+$ in \eqref{control_i1} satisfy
\begin{align}
\label{convergence_condition1}
k_4-s_1k_2-\frac{k_2^2(k_3-s_1k_1)}{k_1k_2-k_4}>0,~\frac{k_1k_2-k_4}{k_2}>0.
\end{align} 
\end{lemma}

{\it Proof.} 
\label{proof_lemma3}
Let $\bar{x},\bar{v}$, be the position, velocity center of the agent set $\mathcal V$ as
\begin{align}
\label{center_state}
 \bar{x}:=\frac{1}{n}\sum_{i=1}^n x_i,  \bar{v}:=\frac{1}{n}\sum_{i=1}^n v_i.
 \end{align}
From the definitions of $x_i,v_i$ in Eq.~\eqref{kinetic_F}, the dynamics of $\bar{x}, \bar{v}$ in \eqref{center_state} are given in the
Cartesian coordinates as follows,
\begin{align}
\label{center_dynamic}
\dot{ \bar{x}}(t)=&\bar{v}(t), ~\dot{ \bar{v}}(t)=\bar{u}(t),
\end{align}
where $\bar{u}=1/n\sum_{i=1}^n u_i$ is the acceleration of center of all the agents. By the definition of $\eta_i$ in Eq.~\eqref{neat_eta}, one has $\sum_{i=1}^n\eta_i=0$, which leads to that
\begin{align}
\label{center_input1}
\bar{u}
        =&-\Big(\begin{bmatrix}
        k_1 & k_2
        \end{bmatrix}\otimes I_2\Big)
        \begin{bmatrix}
        \bar{x}\\
        \bar{v}\\
        \end{bmatrix}-
        \Big(
        \begin{bmatrix}
        k_3 & k_4
        \end{bmatrix}\otimes I_2\Big)
        \begin{bmatrix}
        \bar{\epsilon}\\
        \bar{\zeta}\\
        \end{bmatrix}
\end{align}
with $\bar{\epsilon}:=1/n\sum_{i=1}^n \epsilon_i,\bar{\zeta}:=1/n\sum_{i=1}^n \zeta_i$ being the states of the internal model for the center of agents. Denoting $\varsigma:=[\bar{x}\t, \bar{v}\t]\t$ and $\chi:=[\bar{\epsilon}\t, \bar{\zeta}\t]\t$, the dynamic of the center of agents has the following structure by substituting Eq.~\eqref{center_input1} into Eq.~\eqref{center_dynamic} 
\begin{align}
\label{com_varsigma}
\dot{\varsigma}=& A\varsigma+BK_2\chi,
\end{align}
with $K_2=[-k_3 ~~-k_4]\otimes I_2$, $B=[0~~ 1]\t\otimes I_2$
\begin{align}
\label{matrix_AB}
A=&\begin{bmatrix}
			0 & 1\\
			-k_1 & -k_2
    \end{bmatrix}\otimes I_2.
\end{align}
Combining with the fact $k_5\sum_{i=1}^n \eta_i=0$ and 
$\chi=[\bar{\epsilon}\t, \bar{\zeta}\t]\t$ in Eq.~(\ref{com_varsigma}), it then follows from~\eqref{center_input1} and internal model in \eqref{control_i1} that the dynamics of $\chi$ are calculated as 
\begin{align}
\label{varsigma_d}
\dot{\chi}=&G_1\chi+G_2\bar{e}
\end{align}
with $\bar{e}:=1/n\sum_{i=1}^n e_i$ being the center of the relative position error and 
\begin{align}
\label{matrix_G1G2}
G_1= \begin{bmatrix}
                             0 & 1\\
                             s_1 & 0\\                             
          \end{bmatrix}\otimes I_2,
G_2= \begin{bmatrix}
                             0 \\
                             1 \\                             
          \end{bmatrix}\otimes I_2.        
\end{align}
Denote $\sigma:=[x_d\t, v_d\t]\t$, and the dynamic of the target (see Eq.~\eqref{kinetic_L}) can be rewritten in a compact form as
\begin{align}
\label{target_compect}
\dot{\sigma}=S\sigma.
\end{align} 
Recalling the definitions of $e_i:=x_i-x_d$ and $\varsigma:=[\bar{x}\t, \bar{v}\t]\t$, one has 
\begin{align}
\label{center_e}
\bar{e}
	  =\bar{x}-x_d= C\varsigma+D\sigma
\end{align}
with
\begin{align}
\label{matrix_CD}
C=\begin{bmatrix}
	  1 & 0
	  \end{bmatrix}\otimes I_2,
	  D=\begin{bmatrix}
	  -1 & 0
	  \end{bmatrix}\otimes I_2.
\end{align}
Denoting $\Phi:=[\varsigma\t, \chi\t]\t$, it follows from Eqs.~\eqref{com_varsigma},~\eqref{varsigma_d},
~\eqref{target_compect},~\eqref{center_e} that the augmented system of the center states is
\begin{align}
\label{closed-system}
\dot{\Phi}=& A_c\Phi+B_c\sigma,~\dot{\sigma}=S\sigma,~\bar{e}=C_c{\Phi}+D\sigma
\end{align} 
with 
\begin{align*}
A_c=
\begin{bmatrix}
A & BK_2\\
G_2C & G_1
\end{bmatrix},
B_c=
\begin{bmatrix}
\bf{0}\\
G_2D
\end{bmatrix},
C_c=
\begin{bmatrix}
C & \bf{0}
\end{bmatrix}.
\end{align*}
From the definitions of matrices $A, B, K_2, C, D, G_1, G_2$ in~\eqref{com_varsigma},~\eqref{matrix_G1G2},~\eqref{matrix_CD}, one has
 \begin{align}
\label{matrix_AcBc}
A_c=\begin{bmatrix}
0 & 1 & 0 & 0\\
-k_1 & -k_2 & -k_3 & -k_4\\
0 & 0 & 0 & 1\\
1& 0 & s_1 & 0\
\end{bmatrix}\otimes I_2, 
B_c=\begin{bmatrix}
0 & 0\\
0 & 0\\
0 & 0\\
-1 & 0\\
\end{bmatrix}\otimes I_2.
\end{align}
Then, the characteristic polynomial of matrix $A_c$ is $\lambda^4+k_2\lambda_2^3+(k_1-s_1)\lambda^2+(k_4-s_1k_2)\lambda+k_3-s_1k_1=0.$
Since $s_1\leq0$, one has all coefficients of the polynomial satisfying $k_2>0, k_1-s_1>0, k_4-s_1k_2>0, k_3-s_1k_1>0$, which implies that $A_c$ is Hurwitz iff
\begin{align*}
k_4-s_1k_2-\frac{k_2^2(k_3-s_1k_1)}{k_1k_2-k_4}>0,~\frac{k_1k_2-k_4}{k_2}>0
\end{align*} 
with Routh stability criterion \cite{routh1877treatise}.
Recalling the dynamic matrix $S$ in Eq.~\eqref{matrix_target}, one has that the minimal characteristic polynomial of $S(\lambda)$ is calculated as $\lambda^2-s_1=0.$
Since the regulated output $\bar{e}:=\bar{x}-x_d$ in \eqref{varsigma_d}, one has that $p=1$. Moreover, by Lemma~\ref{p_copy_model}, one has that 
\begin{align*}
\beta_1= \begin{bmatrix}
                             0 & 1\\
                             s_1 & 0\\                             
          \end{bmatrix},
\sigma_1= \begin{bmatrix}
                             0 \\
                             1 \\                             
          \end{bmatrix},        
\end{align*}
which satisfies the conditions (1) and (2) of the minimal $p$-copy internal model in Lemma~\ref{p_copy_model}. Since $\bar{x}\in\mathbb{R}^2, x_d\in\mathbb{R}^2$, the matrices $G_1, G_2$ of the internal model are thus expanded by Kronecker product as follows 
\begin{align*}
G_1= \begin{bmatrix}
                             0 & 1\\
                             s_1 & 0\\                             
          \end{bmatrix}\otimes I_2,
G_2= \begin{bmatrix}
                             0 \\
                             1 \\                             
          \end{bmatrix}\otimes I_2,        
\end{align*}
which is consistent with Eq.~\eqref{matrix_G1G2}. It implies that
$G_1, G_2$ in \eqref{matrix_G1G2} incorporate a minimal p-copy internal model of the matrix $S$ \cite{huang2004nonlinear}. Then, by Lemma~\ref{lemma_ORP}, the closed-loop system in~\eqref{closed-system} satisfies a Sylvester equation with a unique matrix $\mathcal X_c\in \mathbb{R}^{4\times4}$ below
\begin{align}
\label{output_equation1}
\mathcal X_c S=A_c\mathcal X_c+B_c,~\mathbf{0}=C_c\mathcal X_c+D
\end{align}
with $A_c, B_c, C_c, D$ given in \eqref{closed-system}. Let $\widetilde{\Phi}:=\Phi-\mathcal X_c\sigma$ be the errors between the center states $\Phi$ and the solution states $\mathcal X_c\sigma$, it follows from Eqs.~\eqref{closed-system} and \eqref{output_equation1} that $\dot{\widetilde{\Phi}}=A_c\widetilde{\Phi}, \bar{e}=C_c\widetilde{\Phi}$,
which implies that $\lim_{t\rightarrow\infty}\widetilde{\Phi}(t)=0,\lim_{t\rightarrow\infty}\bar{e}(t)=0$ because of $A_c$ is Hurwitz.  

Bearing in mind of~\eqref{center_state} and~\eqref{center_e}, one has $\lim_{t\rightarrow\infty}\bar{x}(t)-x_d(t)={1}/{n}\sum_{i=1}^n x_i(t)-x_d(t)=\bar{e}(t)=0$,
which thus proves that $\lim_{t\rightarrow\infty}P_{x_d(t)}(x(t)) = 0$, i.e., the fencing property in P1.  
\eproof

\begin{remark}
\label{remark_fencing}
The condition in \eqref{convergence_condition1} only guarantees the fencing property of an MAS $\mathcal V$. Moreover, 
this condition~\eqref{convergence_condition1} can be satisfied with the condition~C2, which will be proved in Theorem~\ref{controlaw1}.
\end{remark}


\begin{lemma}
\label{collision_aviodance_property}
Under conditions C1, C2, an MAS governed by \eqref{kinetic_F}, \eqref{control_i1} guarantees inter-agent collision avoidance, i.e.,   
$\|x_{i,k}(t)\|>r, \forall t\geq 0, \; \forall i\neq k \in \cal V$.
\end{lemma}

{\it Proof.} 
\label{proof_lemma_4}
Define $\Phi_i:=[x_i\t, v_i\t, \epsilon_i\t, \zeta_i\t]\t$, and it follows from Eqs.~\eqref{kinetic_F}, \eqref{kinetic_L}, \eqref{control_i1}, \eqref{neat_eta}, \eqref{closed-system} that the derivative of $\Phi_i$ becomes
\begin{align}
\label{varsigma_agent}
\dot{\Phi}_i=& A_c\Phi_i+B_c\sigma+k_5E\eta_i
\end{align}
with $E=[0,1,0,-1]\t\otimes I_2\in\mathbb{R}^{8\times2}$, $A_c, B_c$ in \eqref{matrix_AcBc} and $\eta_i$ given in \eqref{neat_eta}.
Let $\widetilde{\Phi}_i:=[\widetilde{x}_i\t, \widetilde{v}_i\t, \widetilde{\epsilon}_i\t, \widetilde{\zeta}_i\t]\t$ be the fencing error as
$\widetilde{\Phi}_i:=\Phi_i-\mathcal X_c\sigma$,
the time derivative of $\widetilde{\Phi}_i$ along  the dynamics~\eqref{closed-system} and~\eqref{varsigma_agent} is
\begin{align*}
\dot{\widetilde{\Phi}}_i
                                    =A_c\widetilde{\Phi}_i+(A_c\mathcal X_c+B_c-\mathcal X_cS)\sigma+k_5E\eta_i,
\end{align*}
which follows from Eq.~\eqref{output_equation1} that
\begin{align}
\label{err_varsigma1}
\dot{\widetilde{\Phi}}_i=&A_c\widetilde{\Phi}_i+k_5E\eta_i.
\end{align}
It is observed from Eq.~\eqref{err_varsigma1} that the inter-agent repulsion $\eta_i$ of agent $i$ exists in the dynamic of $\widetilde{\Phi}_i$, which influences the convergence of the closed-loop error system~\eqref{err_varsigma1}.

Moreover, the integration of inter-agent
repulsion $\sum_{i\in\mathcal V}\sum_{k\in\mathcal N_i}\int_{\|x_{i,k}\|}^{R}\alpha(\tau)d\tau$ is introduced as a potential function in a Lyapunov candidate to guarantee inter-agent collision avoidance. Denoting 
\begin{align}
\label{V_p}
V_p:=\sum_{i\in\mathcal V}\sum_{k\in\mathcal N_i}\int_{\|x_{i,k}\|}^{R}\alpha(\tau)d\tau
\end{align} 
as the integration of inter-agent
repulsion for conciseness, one has that the derivative of $V_p$ wirtes 
\begin{align}
\label{derivative_repulsion}
\frac{dV_p}{dt}
=&-\sum_{i\in\mathcal V}\sum_{k\in\mathcal N_i}\alpha(\|x_{i,k}\|)\frac{d\|x_{i,k}\|}{dt}\nonumber\\
=&-\sum_{i\in\mathcal V}\sum_{k\in\mathcal N_i}\alpha(\|x_{i,k}\|)\frac{x_{i,k}\t}{\|x_{i,k}\|}(\frac{\partial x_{i,k}}{\partial x_i}\dot{x}_i+\frac{\partial x_{i,k}}{\partial x_i}\dot{x}_k)\nonumber\\
=&-2\sum_{i\in\mathcal V}\sum_{k\in\mathcal N_i}\alpha(\|x_{i,k}\|)\frac{x_{i,k}\t}{\|x_{i,k}\|}\dot{x}_i.
\end{align}
Using the definition of $\eta_i$ in Eq.~\eqref{neat_eta}, one has that $\sum_{i\in\mathcal V}\eta_i\t v_d=v_d\t\sum_{i\in\mathcal V}\eta_i=0$. Then it follows from Eq.~\eqref{derivative_repulsion} that
\begin{align}
\label{dV_part1}
\frac{dV_p}{dt}=&-2\sum_{i\in\mathcal V}\eta_i\t\dot{x}_{i}=-2\sum_{i\in\mathcal V}\eta_i\t(\dot{x}_{i}-v_d)=-2\sum_{i\in\mathcal V}\eta_i\t\widetilde{v}_{i}
\end{align}
with $\dot{x}_i=v_i$ and $\widetilde{v}_i:=v_i-v_d$. It can be deduced that $\eta_i$ is coupled with error state $\widetilde{v}_{i}$ in Eq.~\eqref{dV_part1}.

Construct a Lyapunov candidate consisting of the error states $\widetilde{\Phi}_i$ and the potential function $V_p$ as follows, 
\begin{align}
\label{V1}
V_1(\widetilde{\Phi}_i, x_{i,k})=&\sum_{i\in\mathcal V}\Big\{\widetilde{\Phi}_i\t P\widetilde{\Phi}_i\Big\}+\gamma k_5V_p,
\end{align}
where $P\in\mathbb{R}^{8\times 8}$ is a positive-definite symmetrical matrix, and $\gamma$ is the parameter associated with the matrix $P$. 
Then, from the derivative of $V_p$ in \eqref{dV_part1}, the time derivative of $V_1(\widetilde{\Phi}_i, x_{i,k})$, along the trajectories of~\eqref{err_varsigma1} becomes
\begin{align}
\label{dot_V1}
\frac{dV_1(\widetilde{\Phi}_i, x_{i,k})}{dt}
                             			 =&\sum_{i\in\mathcal V}\Big\{\widetilde{\Phi}_i\t (PA_c +A_c\t P)\widetilde{\Phi}_i\nonumber\\
					      	   &+k_5\Big(\widetilde{\Phi}_i\t PE\eta_i+\eta_i\t E\t P\widetilde{\Phi}_i\Big)\Big\}\nonumber\\
						   &-2\gamma k_5\sum_{i\in\mathcal V}\eta_i\t\widetilde{v}_{i}
.
\end{align}
In order to rewrite the term of $\widetilde{\Phi}_i\t PE\eta_i+\eta_i\t E\t P\widetilde{\Phi}_i$ in the right-hand side of Eq.~\eqref{dot_V1} into the form of $\eta_i\t\widetilde{v}_i$ which can be eliminated by the derivative of $V_p$ in Eq.~\eqref{dV_part1}, the matrix $P$ is designed below,
\begin{align}
\label{P_matrix}
P=
\begin{bmatrix}
p_1 & p_6 & p_7 & p_6\\
p_6 & p_2 & p_5 & p_4\\
p_7 & p_5 & p_3 & p_5\\
p_6 & p_4 & p_5 & p_4\\
\end{bmatrix}\otimes I_2
\end{align}
with parameters $p_i, i=1,\cdots, 7,$ to be designed later. Comparing matrices $E, P$ in Eqs.~\eqref{varsigma_agent} and~\eqref{P_matrix} yields
\begin{align}
\label{dV_part2}
k_5\Big(\widetilde{\Phi}_i\t PE\eta_i+\eta_i\t E\t P\widetilde{\Phi}_i\Big)=2k_5(p_2-p_4) \widetilde{v}_i\eta_i,
\end{align}
which implies that the parameter $\gamma$ in Eq.~\eqref{V1} can be designed to be $\gamma=p_2-p_4$.

Accordingly, it follows from Eqs.~\eqref{dot_V1}, \eqref{dV_part1}, \eqref{dV_part2} that
\begin{align}
\label{dot_V2}
\frac{dV_1(\widetilde{\Phi}_i, x_{i,k})}{dt}
						 =&\sum_{i\in\mathcal V}\Big\{\widetilde{\Phi}_i\t Q\widetilde{\Phi}_i\Big\}
\end{align}
with
\begin{align}
\label{Q_matrix}
Q=&
\begin{bmatrix}
q_1 & q_5 & q_6 & q_7\\
q_5 & q_2 & q_8 & q_{9}\\
q_6 & q_8 & q_3 & q_{10}\\
q_7 & q_9 & q_{10} & q_4\\
\end{bmatrix}\otimes I_2,
\end{align}
$q_1=-2k_1p_6+2p_6,~q_2=2p_6-2k_2p_2,
q_3=-2k_3p_5+2s_1p_5,~q_4=-2k_4p_4+2p_5,
q_5=p_1-k_2p_6-k_1p_2+p_4,
q_6=-k_3p_6+s_1p_6-k_1p_5+p_5,
q_7=p_4+p_7-k_4p_6-k_1p_4,
q_8=-k_3p_2+s_1p_4+p_7-k_2p_5,
q_9=-k_4p_2+p_5+p_6-k_2p_4,
q_{10}=-k_4p_5+p_3-k_3p_4+s_1p_4$.

To prove the convergence of $\widetilde{v}_i$ and $\widetilde{\zeta}_i$ in Eq.~\eqref{dot_V2}, $q_i, i=1, 3, 5, 6, 7, 8, 10,$ in \eqref{Q_matrix} can be set as $0$, which implies that 
$p_5=0, p_6=0$, $p_1-k_1p_2+p_4=0, p_4+p_7-k_1p_4=0, -k_3p_2+s_1p_4+p_7=0$ and $p_3-k_3p_4+s_1p_4=0$.
Then it derives that
\begin{align}
\label{parameter_p1}
p_1=&\frac{k_1^2+(s_1-1)k_1-k_3}{k_3}p_4,~p_2=\frac{k_1+s_1-1}{k_3}p_4,\nonumber\\
p_3=&(k_3-s_1)p_4,~p_7=(k_1-1)p_4=(k_3-s_1)p_2, 
\end{align}
which implies that 
\begin{align}
\label{P_matrix1}
P=
\begin{bmatrix}
\frac{k_1^2+(s_1-1)k_1-k_3}{k_3} & 0 & k_1-1 & 0\\
0 & \frac{k_1+s_1-1}{k_3} & 0 & 1\\
k_1-1 & 0 & k_3-s_1 & 0\\
0 & 1 & 0 & 1\\
\end{bmatrix}p_4\otimes I_2.
\end{align}
It follows from $k_3>0, s_1\leq0, p_4>0$ that the last two diagonal term of $P$ satisfy $k_3-s_1>0$ and $1>0$. 
Then, in accordance to Lemma~\ref{lemma_schur}, one has
\begin{align}
\label{schur_inequality}
&\begin{bmatrix}
\frac{k_1^2+(s_1-1)k_1-k_3}{k_3} & 0 \\
0 & \frac{k_1+s_1-1}{k_3}\\
\end{bmatrix}-\begin{bmatrix}
\frac{(k_1-1)^2}{k_3-s_1}& 0 \\
0 & 1\\
\end{bmatrix}>0
\end{align}
if $P>0$. Direct calculation of \eqref{schur_inequality} gives 
\begin{align}
\label{schur_inequality1}
\begin{bmatrix}
\xi_1 & 0 \\
0 &\xi_2\\
\end{bmatrix}>0
\end{align}
with $\xi_1:=\big\{(s_1+1)k_1k_3-k_3^2-s_1k_1^2-(s_1-1)s_1k_1+(s_1-1)k_3)\big\}/(k_3^2-k_3s_1)$ and $\xi_2:=(k_1-k_3+s_1-1)/k_3.$
As $k_3>0, k_3-s_1>0$, we only consider the numerators of $\xi_1,\xi_2$ to determine $P>0$. 

Case 1: If $s_1=0$, it is derived that $(k_1-k_3+s_1-1)/k_3>0$ for $P>0$. 

Case 2: If $s_1<0$, rewrite the numerator of $\xi_1$ in~\eqref{schur_inequality1} in descending $k_1$ order as $f(k_1)=-s_1k_1^2+\big((s_1+1)k_3-(s_1-1)s_1\big)k_1-k_3^2+(s_1-1)k_3$,
which implies that $f(k_1)$ is a quadratic function of $k_1$.
From the fact that 
\begin{align*}
&k_3+1-s_1-\frac{(s_1+1)k_3-(s_1-1)s_1}{2s_1}\nonumber\\
=&\frac{(s_1-1)(k_3-s_1)}{2s_1}>0,
\end{align*}
one has that $f(k_1)$ monotonically increases when $k_1>k_3+1-s_1$. 
Substituting $k_1=k_3+1-s_1$ into $f(k_1)$ yields that 
$f(k_1)=-s_1(k_3+1-s_1)^2+\big((s_1+1)k_3-(s_1-1)s_1\big)(k_3+1-s_1)-k_3^2+(s_1-1)k_3=0$,
which implies that $f(k_1)>0$ (i.e., $P>0$) if $k_1>k_3+1-s_1$.  

Accordingly, it concludes that $\xi_1>0, \xi_2>0$ (i.e., $P>0$) if $(k_1-k_3+s_1-1)/k_3>0$. Note that $k_1^2+(s_1-1)k_1-k_3=k_1(k_1+s_1-1)-k_3>0$ with $k_1>1$ and $k_1-1-k_3+s_1>0$ in~\eqref{P_matrix1}. 
Moreover, the condition $k_1+s_1-1>k_3$ implies $p_2-p_4>0$ with \eqref{parameter_p1}, which guarantees $V_1$ in Eq.~\eqref{V1} is positive definite.

Next, it follows from the fact $p_5=0, p_6=0$ and Eq.~\eqref{parameter_p1} that
\begin{align}
\label{dot_V3}
\frac{dV_1(\widetilde{\Phi}_i, x_{i,k})}{dt}
						 =&\sum_{i\in\mathcal V}\Bigg\{
						 \begin{bmatrix}
						  \widetilde{v}_i\\
						  \widetilde{\zeta}_i
						 \end{bmatrix}\t						 
						 \bar{Q}
						 \begin{bmatrix}
						  \widetilde{v}_i\\
						  \widetilde{\zeta}_i
						 \end{bmatrix}
						 \Bigg\}
\end{align}
with 
\begin{align*}
\bar{Q}=
 \begin{bmatrix}
						 -2k_2p_2 & -k_4p_2-k_2p_4\\
						 -k_4p_2-k_2p_4 & -2k_4p_4
						 \end{bmatrix}\otimes I_2.
\end{align*}
Due to the fact that the leading principal minors of $\bar{Q}$ fulfilling $-2k_2p_2<0$ and $-2k_2p_2\times-2k_4p_4-(-k_4p_2-k_2p_4)^2=-(k_4p_2-k_2p_4)^2$.
It follows from Eq.~\eqref{parameter_p1} that $k_4p_2-k_2p_4=0$ if
\begin{align}
\label{equality_parameter}
k_4\frac{k_1+s_1-1}{k_3}=k_2,
\end{align}  
which then  implies that the matrix $\bar{Q}$ is negative semidefinite.
Combining the condition $k_1>k_3+1-s_1$ and Eq.~\eqref{equality_parameter} together gives  C2.
Moreover, one has that 
\begin{align}
\label{dot_V4}
\frac{dV_1(\widetilde{\Phi}_i, x_{i,k})}{dt}
						 =&-\sum_{i\in\mathcal V}\frac{2p_4}{k_4}\|(k_2						   		
						   \widetilde{v}_i+k_4\widetilde{\zeta}_i)\|^2\leq0.
\end{align}
Denoting $V_1(t):=V_1(\widetilde{\Phi}_i(t), x_{i,k}(t)), \forall t\geq0$ as the function of $V_1$ at time $t$ for conciseness, it then follows from Eq.~\eqref{dot_V4} that 
\begin{align}
\label{Integre_V}
V_1(\overline{T})=& V_1(0)+\int_0^{\overline{T}} \frac{dV_1(\widetilde{\Phi}_i, x_{i,k})}{dt}dt\leq V_1(0)
\end{align}
for an arbitrary constant time $\overline{T}>0$. Combining with the definition of $V_1$ in Eq.~\eqref{V1}, one has that 
\begin{align}
\label{bounded_potential}
&k_5(p_2-p_4)\sum_{i\in\mathcal V}\sum_{k\in\mathcal N_i}\int_{\|x_{i,k}(\overline{T})\|}^{R}\alpha(s)ds\nonumber\\
\leq& V_1(\overline{T})\leq V_1(0).
\end{align}
Under condition C1, one has that $V_1(0)$ is bounded, so is $\sum_{i\in\mathcal V}\sum_{k\in\mathcal N_i}\int_{\|x_{i,k}(\overline{T})\|}^{R}\alpha(s)ds, \forall~\overline{T}\geq0$ in Eq.~\eqref{bounded_potential}. However, using the fact  
\begin{align}
\lim_{\|x_{i,k}(\overline{T})\|\rightarrow r^{+}}&\sum_{i\in\mathcal V}\sum_{k\in\mathcal N_i}\int_{\|x_{i,k}(\overline{T})\|}^{R}\alpha(s)ds=\infty, 
\end{align}
with $\alpha$ given in \eqref{form_potentional_function} and $r^+$ the right limit of $r$,   
we conclude that $\|x_{i,k}(\overline{T})\|>r, \forall i\in \cal V, i, k \in \mathcal V, \overline{T}>0$. 
Collision avoidance is thus proved. 
\eproof

\begin{remark}
Lemma~\ref{collision_aviodance_property} is to prove inter-agent collision avoidance via the boundness of the Lyaponov function~$V_1$ in~\eqref{V1}, which
consists of two terms. The first term concerns the error states $\widetilde{\Phi}_i$ between the agent $i$ and the target, whereas the second term the integration of inter-agent repulsion $V_p$. Essentially, the aforementioned two terms in $V_1$ can not converge to zero, but only to an invariant set of partial states $k_2\widetilde{v}_i+k_4\widetilde{\zeta}_i=0$ because of~\eqref{dot_V4}.  
Moreover, if the above two terms converge to zeros, it implies that the positions of all the agents will coincide together and thus $V_1=\infty$, which then contradicts the condition of $dV_1(t)/dt\leq0$ in \eqref{dot_V4}. It is still worth mentioning that the selected function $V_1$ and the invariant set $k_2\widetilde{v}_i+k_4\widetilde{\zeta}_i=0$ also contribute to the rigid formation in Lemma~\ref{rigid_property}.  
\end{remark}

\begin{lemma}
\label{rigid_property}
Under conditions C1-C3, an MAS governed by \eqref{kinetic_F}, \eqref{control_i1} achieves a rigid formation moving with the target \eqref{kinetic_L}, i.e., $\lim_{t\rightarrow\infty}v_i(t)-v_d(t)=0$. 
\end{lemma}

{\it Proof.}
\label{proof_lemma_5}
Recalling Eq.~\eqref{dot_V4} in the proof of Lemma~\ref{collision_aviodance_property},  ${dV_1}/{dt}=0$ only if $k_2\widetilde{v}_i+k_4\widetilde{\zeta}_i=0$, which implies that the largest invariant set 
$\{\widetilde{x}_i,\widetilde{v}_i, \widetilde{\epsilon}_i, \widetilde{\zeta}_i~|~{dV_1}/{dt}=0\}$ only contains a line set $\{k_2\widetilde{v}_i+k_4\widetilde{\zeta}_i=0\}$.
Since the line segment of any two nodes from the line set is still in the line set, the line invariant set is thus compact. In accordance to the Lasalle invariant set theorem~\cite{khalil2002nonlinear}, the trajectories of $\widetilde{x}_i,\widetilde{v}_i, \widetilde{\epsilon}_i, \widetilde{\zeta}_i$
converge to 
 \begin{align}
 \label{limit_set}
\lim_{t\rightarrow\infty}k_2\widetilde{v}_i(t)+k_4\widetilde{\zeta}_i(t)=0.
 \end{align}
 Next, we will prove the statement that
 \begin{align}
 \label{equal_relation}
& \mbox{If}~k_2\widetilde{v}_i+k_4\widetilde{\zeta}_i=0,~\mbox{then}~v_i(t)-v_d(t)=0, \forall i\in \cal V.
 \end{align}
According to the dynamics of $\dot{\widetilde{v}}_i,\dot{\widetilde{\zeta}}_i$ in \eqref{err_varsigma1}, one has $\dot{\widetilde{\zeta}_i}=\widetilde{x}_i+s_1\widetilde{\epsilon}_i-k_5\eta_i$.
Taking the derivative of $\dot{\widetilde{v}}_i$ yields that
\begin{align}
\label{value:widetilde_v1}
\ddot{\widetilde{v}}_i
			       =&-(k_2\dot{\widetilde{v}}_i+k_4\dot{\widetilde{\zeta}}_i)-\bigg(k_1-1-\frac{(k_3-s_1)k_2}{k_4}\bigg)\widetilde{v}_i\nonumber\\
			       &-(k_3-s_1)\bigg(\frac{k_2}{k_4}\widetilde{v}_i+\widetilde{\zeta}_i\bigg)-\ddot{\widetilde{\zeta}}_i.
\end{align}
Subtracting ${k_2}/{k_4}\ddot{\widetilde{v}}_i$ at both sides of \eqref{value:widetilde_v1}, it then follows from Eq.~\eqref{equality_parameter} that
\begin{align}
\label{value:widetilde_v2}
 \ddot{\widetilde{v}}_i
			       =&s_1\widetilde{v}_i+\frac{k_4}{k_2-k_4}(k_2\dot{\widetilde{v}}_i+k_4\dot{\widetilde{\zeta}}_i)+\frac{(k_3-s_1)}{k_2-k_4}\nonumber\\
			       &({k_2}\widetilde{v}_i+k_4\widetilde{\zeta}_i)+\frac{1}{k_2-k_4}({k_2}\ddot{\widetilde{v}}_i+k_4\ddot{\widetilde{\zeta}}_i).
\end{align}
By the virtue of uniformly continuous $\dot{\widetilde{v}}_i,\dot{\widetilde{\zeta}}_i, \ddot{\widetilde{v}}_i, \ddot{\widetilde{\zeta}}_i$, it can be deduced from \eqref{equal_relation} that 
\begin{align}
\label{invariant_set1}
k_2\dot{\widetilde{v}}_i+k_4\dot{\widetilde{\zeta}}_i=0,~k_2\ddot{\widetilde{v}}_i+k_4\ddot{\widetilde{\zeta}}_i=0.
\end{align}
Comparing \eqref{equal_relation}, \eqref{value:widetilde_v2}, \eqref{invariant_set1} gives 
\begin{align}
\label{value:widetilde_v3}
\ddot{\widetilde{v}}_i
			       =&s_1\widetilde{v}_i,
\end{align}
which implies that $\ddot{\widetilde{\zeta}}_i=s_1\widetilde{\zeta}_i$.
Since $\dot{\widetilde{\epsilon}}_i=\widetilde{\zeta}_i$, one has that
 \begin{align}
\label{value:widetilde_v4}
\dot{\widetilde{\zeta}}_i(t)=s_1\widetilde{\epsilon}_i(t)-m_i
\end{align}
with a proper constant vector $m_i\in\mathbb{R}^2$. It, together with the fact $\dot{\widetilde{\zeta}_i}=
s_1\widetilde{\epsilon}_i(t)+\widetilde{x}_i-k_5\eta_i$ in \eqref{err_varsigma1}, gives
\begin{align}
\label{value:widetilde_v8}
\widetilde{x}_i(t)-k_5\sum_{k\in\mathcal N_i} \bigg\{\alpha(\|x_{i,k}\|)\frac{{x_{i,k}}}{{\|x_{i,k}\|}}\bigg\}=m_i, \forall i\in \cal V.
\end{align}
Case 1: If $s_1=0$, the present problem reduces to a fencing one with a constant-velocity target (see, e.g., \cite{kou2021cooperative}). 
It follows from Eq.~\eqref{value:widetilde_v3} that $\lim_{t\rightarrow\infty}\ddot{\widetilde{v}}_i(t)=0$. Then, it suffices to prove that $\lim_{t\rightarrow\infty}\dot{\widetilde{v}}_i(t)=0, \lim_{t\rightarrow\infty}\widetilde{v}_i(t)=0$ by contradiction, which is similar to \cite{kou2021cooperative}. Hence, it concludes that $ \lim_{t\rightarrow\infty}\widetilde{v}_i(t)=v_i(t)-v_d(t)=0, \forall i\in \cal V$.

Case 2: If $s_1<0$, it follows from Eq.~\eqref{value:widetilde_v3} and differential equation \cite{hormander1967hypoelliptic} that the solution of $\widetilde{v}_i$ is calculated as
\begin{align}
\label{solution_widetilde_v1}
\widetilde{v}_i(t)=c_{i,1}\cos(\sqrt{-s_1}t)+c_{i,2}\sin(\sqrt{-s_1}t)
\end{align}
for $c_{i,1}, c_{i,2}$ being the parameters designed by initial states.
Next, we will prove $c_{i,1}=0, c_{i,2}=0$ by contradiction. 
It follows from Eqs.~\eqref{err_varsigma1}, \eqref{solution_widetilde_v1} that $\widetilde{x}_i$
\begin{align}
\label{solution_widetilde_v4}
\widetilde{x}_i(t)=\frac{c_{i,1}}{\sqrt{-s_1}}\sin(\sqrt{-s_1}t)-\frac{c_{i,2}}{\sqrt{-s_1}}\cos(\sqrt{-s_1}t)+d_i
\end{align} 
with a constant vector  $d_i\in\mathbb{R}^2$. It can be deduced that $\widetilde{x}_i$ in \eqref{solution_widetilde_v4} is a periodic
function if $c_{i,1}\neq0$ or $c_{i,2}\neq0$, which contradicts with the condition of \eqref{value:widetilde_v8}.
Then, it derives that $c_{i,1}=0, c_{i,2}=0, \forall i\in \cal V$, which implies that $\widetilde{v}_i(t)=v_i(t)-v_d(t)=0$ with \eqref{solution_widetilde_v1} (i.e., the proof of \eqref{equal_relation} is thus completed). 

As the term $k_2\widetilde{v}_i+k_4\widetilde{\zeta}_i$ is uniformly continuous, due to~\eqref{equal_relation}, for any $\delta_1>0$, there exists $\delta_2>0$, such that
 \begin{align*}
& \big\|k_2\widetilde{v}_i+k_4\widetilde{\zeta}_i\big\|<\delta_2, ~\forall i\in \cal V,
 \end{align*}
 which leads to the fact that
 \begin{align*}
& \big\|v_i-v_d\big\|<\delta_1, \forall i\in \cal V.
 \end{align*}
Since $ \lim_{t\rightarrow\infty}k_2\widetilde{v}_i(t)+k_4\widetilde{\zeta}_i(t)=0$ in \eqref{limit_set}, there exists a constant $T>0$
such that $\forall t\geq T$, $\|k_2\widetilde{v}_i+k_4\widetilde{\zeta}_i\|<\delta_2,~\forall i\in \cal V$, which implies that 
$\|v_i-v_d\|<\delta_1,~\forall i\in \cal V$. It thus concludes $\lim_{t\rightarrow\infty}v_i(t)-v_d(t)=0,~\forall i\in \cal V$, i.e., a rigid formation, which completes the proof.
\eproof

%
 
 

With Lemmas~\ref{fenceing_property}-\ref{rigid_property}, it is ready to present the main technical results.
\begin{theorem}
\label{controlaw1}
An MAS $\mathcal V$ composed of \eqref{kinetic_F}, \eqref{control_i1}, and a motional target \eqref{kinetic_L}
achieves the property P1, under the conditions C1, C2 and C3, i.e., Problem~{\it 1} is solved.  
\end{theorem}

{\it Proof.} On one hand, it follows from Lemma~\ref{fenceing_property} that the fencing property is achieved if the control gains $k_1, k_2, k_3, k_4$ satisfy~\eqref{convergence_condition1}. 

On the other hand, in view of collision avoidance and rigid-formation property in Lemmas~\ref{collision_aviodance_property} and \ref{rigid_property}, the control gains $k_1, k_2, k_3, k_4$ are required to 
satisfy another C2 in \eqref{condition_C2}. 
Substituting Eq.~\eqref{condition_C2} into Eq.~\eqref{convergence_condition1} in Lemma~\ref{fenceing_property} yields
\begin{align*}
k_4-s_1k_2-\frac{k_2^2(k_3-s_1k_1)}{k_1k_2-k_4}
						  =\frac{\big(k_1+s_1-1-k_3\big)k_4^2}{k_3(k_1k_2-k_4)}>&0,\nonumber\\
\frac{k_1k_2-k_4}{k_2}
				  =\frac{k_1+s_1-1-k_3}{k_1+s_1-1}>&0,
\end{align*}
which implies that the condition C2 suffices to satisfy Eq.~\eqref{convergence_condition1} (i.e., Remark~\ref{remark_fencing}). It thus concludes Problem 1 is solved with C1-C3. The proof is thus completed.
\eproof

\begin{remark}
Compared with the label-fixed strategies, the label-free design in \eqref{control_i1} can be utilized directly without extra calculation or design upon increasing number of agents. Moreover, the label-free design is more flexible in variational environments and some specific situations, which can take a shortcut to achieve fencing formation. 
Last but not least, the label-free design is more robust upon breaking down of some agents.  
An illustrative example will be given in the next session.
\end{remark}

\section{Numerical Simulation}
Consider an MAS governed by \eqref{kinetic_F}, \eqref{control_i1} with $n=4$ and a moving target (\ref{kinetic_L}) of a variational velocity. The sensing range and safe distance are set to be $R=10, r=2$, respectively, which implies that the potential function $\alpha(\cdot)$ in \eqref{form_potentional_function} is set to be $r=2, R=10$.

\begin{figure}[!htb]
\centering
\includegraphics[width=\hsize]{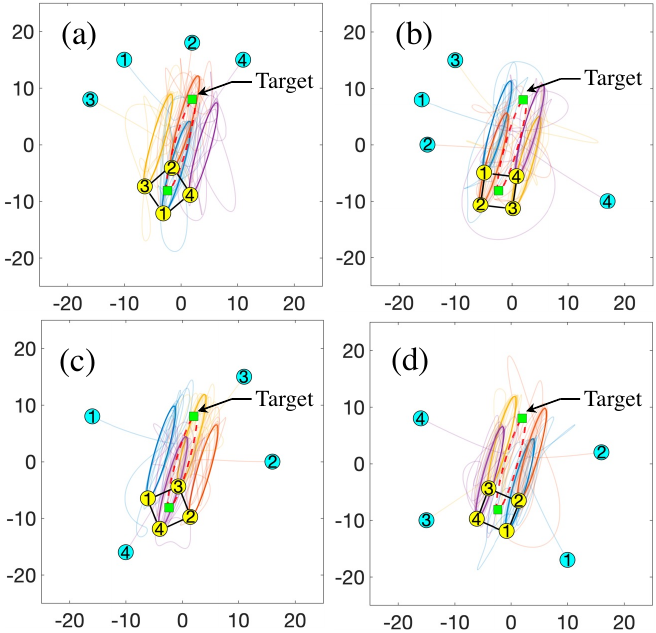}
\caption{(a)-(d) Trajectories of the agents from random initial positions and velocities to collision-free rigid-formation target fencing with the proposed label-free controller~\eqref{control_i1} (Here, the blue and yellow circles denote the initial and final states of the agents, respectively, whereas the green square the target).}
\label{trajectory3}
\end{figure} 
\begin{figure}[!htb]
\centering
\includegraphics[width=\hsize]{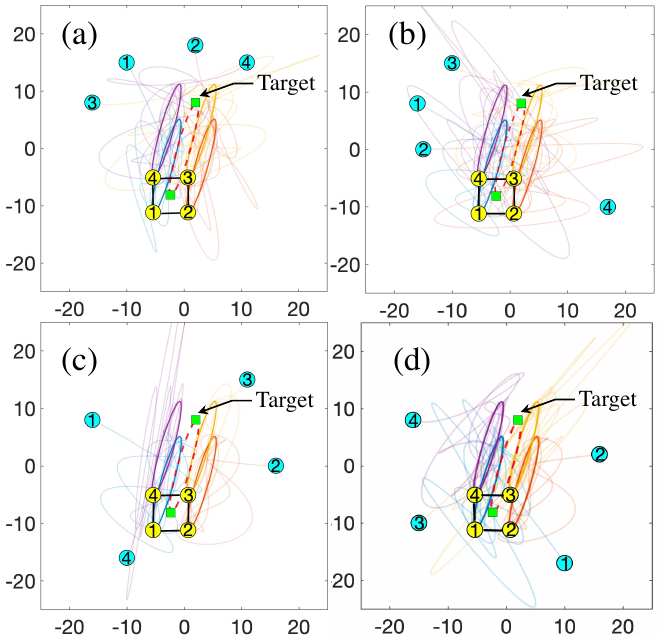}
\caption{ (a)-(d) Trajectories of the agents from the same initial positions and velocities in Fig.~\ref{trajectory3} to collision-free rigid-formation target fencing with a label-fixed strategy (Here, the blue and yellow circles denote the initial and final states of the agents, respectively, whereas the green square the target).}
\label{fixed_trajectory3}
\end{figure} 
   \begin{figure}[!htb]
  \centering
  \includegraphics[width=\hsize]{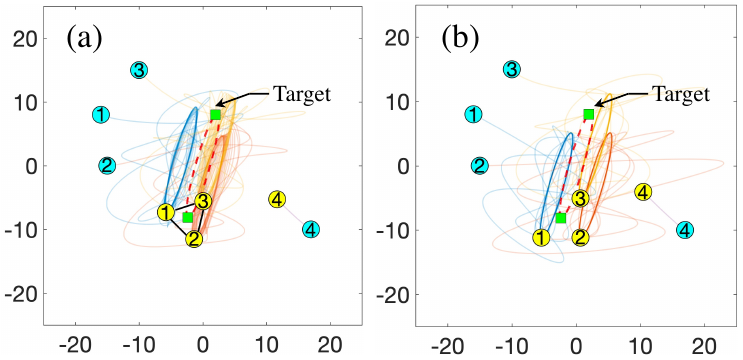}
  \caption{ A special case when agent $4$ suddenly breaks down at $t=20$s. 
  Trajectory comparison of the agents and the target between the proposed label-free fencing controller~\eqref{control_i1} (see subfigure (a)) and a label-fixed strategy (see subfigure (b)) (Here, the blue and yellow circles denote the initial and final states of the agents, respectively whereas the green square the target).}
  \label{sudden_broken3}
\end{figure} 

\begin{figure}[!htb]
  \centering
  \includegraphics[width=8cm]{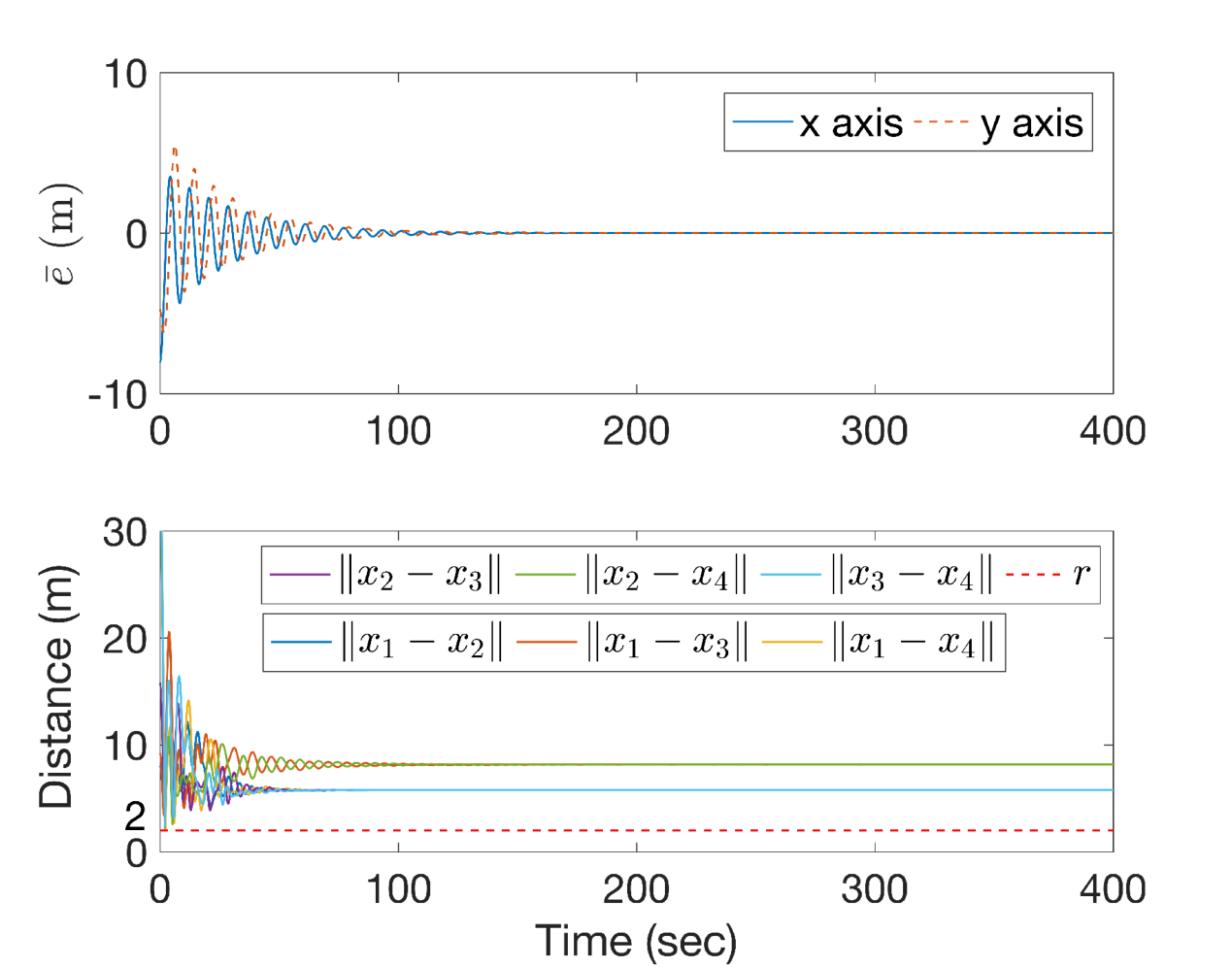}
  \caption{Temporal evolution of the fencing error $\bar{e}$ (see Eq.~\eqref{center_e}) and  the relative distances $\|x_i-x_j\|, i\neq k, i, k \in\cal V$ among agents in Fig.~\ref{trajectory3} (b) for example.}
  \label{distance3}
\end{figure}
\begin{figure}[!htb]
  \centering
  \includegraphics[width=8cm]{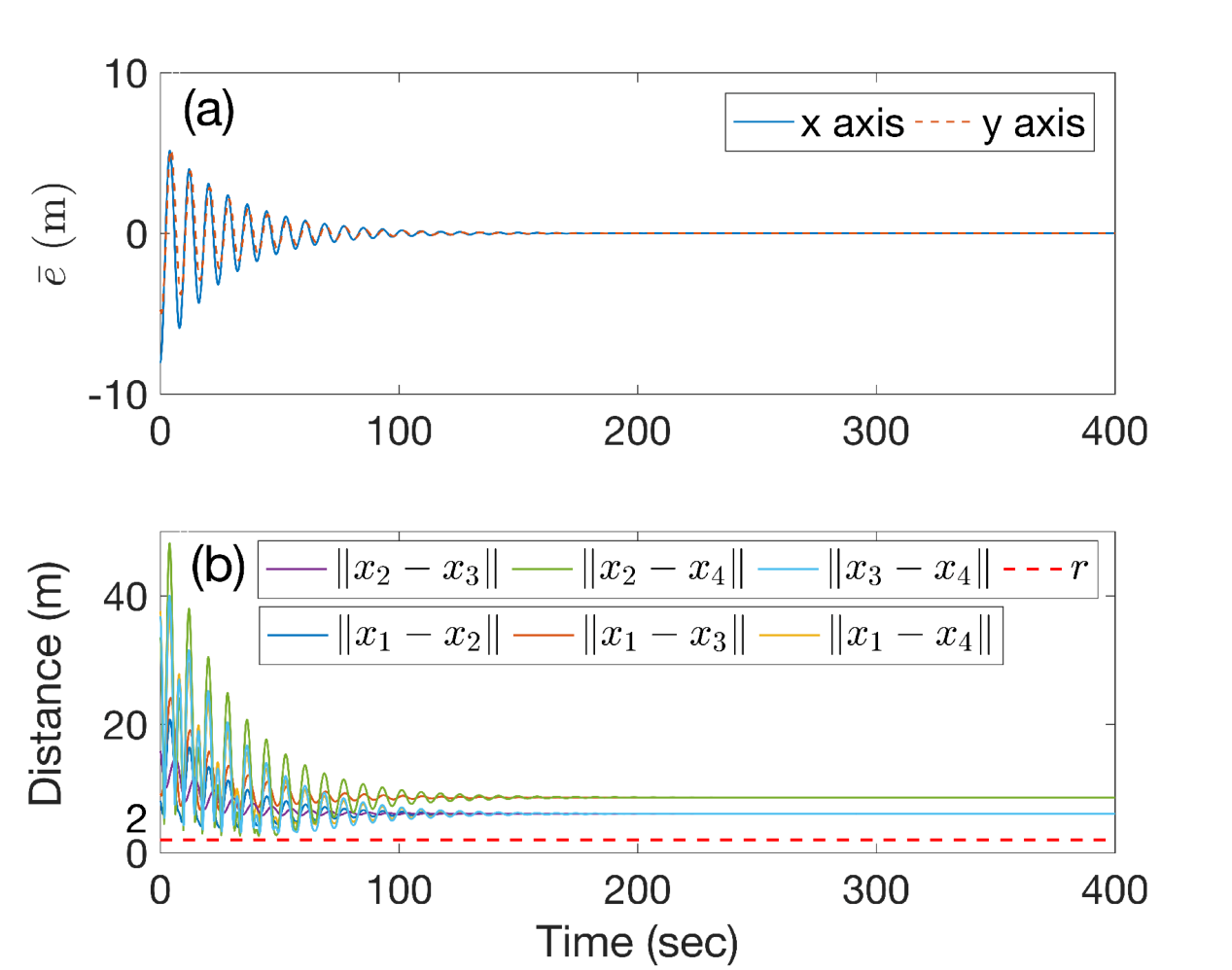}
  \caption{Temporal evolution of the fencing error $\bar{e}$ (see Eq.~\eqref{center_e}) and  the relative distances $\|x_i-x_j\|, i\neq k, i, k \in\cal V$ among agents in Fig.~\ref{fixed_trajectory3} (b) for example.}
  \label{fixed_distance3}
\end{figure}

\begin{figure}[!htb]
  \centering
  \includegraphics[width=8cm]{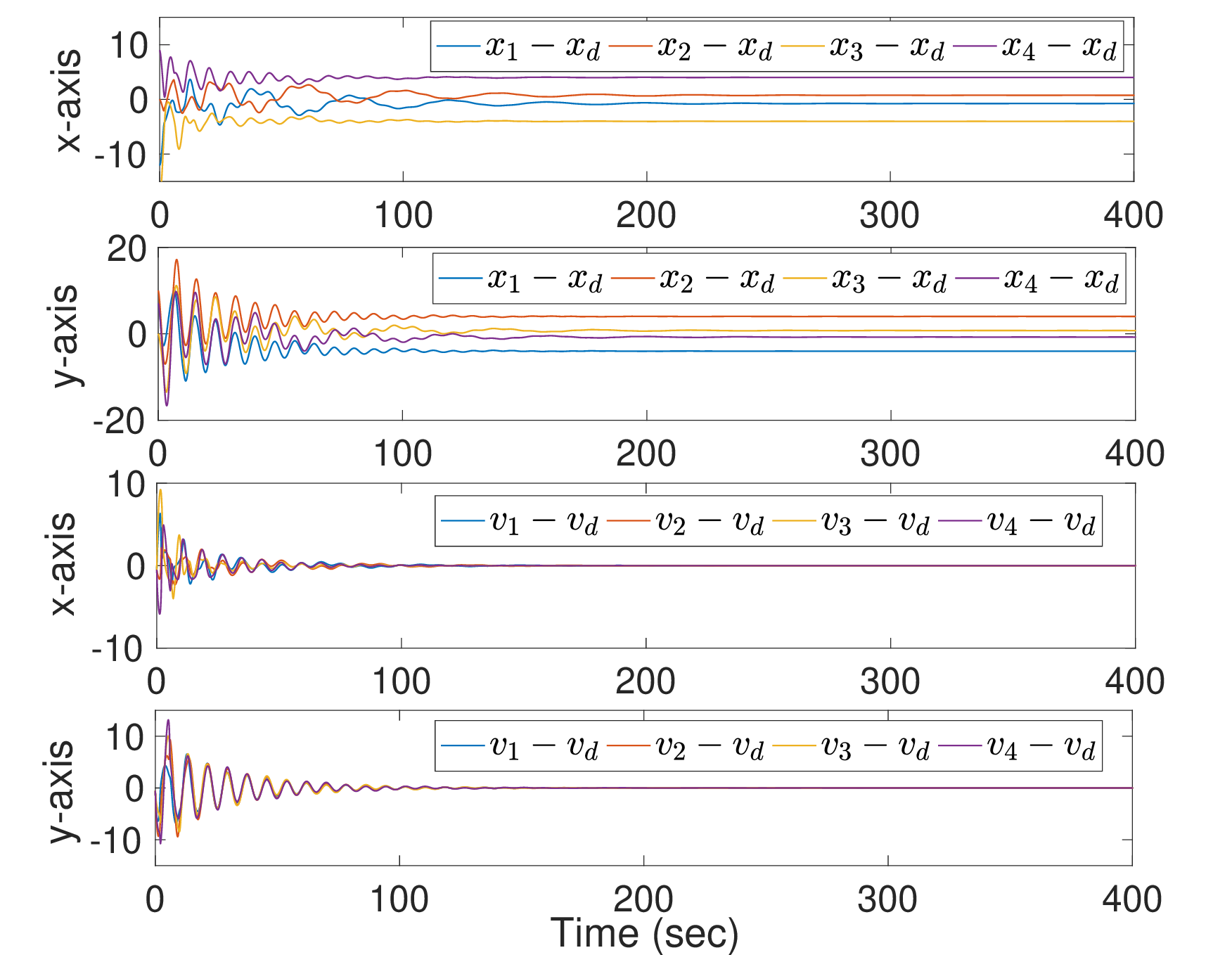}
  \caption{Temporal evolution of the position error $x_i-x_d$ and velocity error $v_i-v_d,  i=1, 2, 3, 4,$ in Fig.~\ref{trajectory3} (b) for example.}
  \label{center3}
\end{figure}

Set $s_1=-0.1$ to assure that the target moves periodically with the initial position $x_d(0)=[2,8]\t$ and the initial velocity $v_d(0)=[0.5, 0.5]\t$, whose moving trajectories are presented by red dashed curves in Fig.~\ref{trajectory3}. The condition C2 is satisfied by picking the parameters in \eqref{control_i1} as $k_1=2.2, k_2=6, k_3=0.1, k_4=3$, and $k_5$ in \eqref{control_i1} is set as $k_5=20$.
Figs.~\ref{trajectory3} (a)-(d) demonstrate the temporal evolution of agents from different initial states (blue circles) fulfilling condition C1 to the final rigid-formation fencing states (yellow circles) with the motional target (green square) satisfying $P_{x_d}(x)=0$. It is observed that the fencing formation of agents from any initial states is achieved with distinct labels. 
To show the advantages of the label-free approach, the corresponding fencing simulations with a label-fixed strategy are conducted in Fig.~\ref{fixed_trajectory3} with the same initial position setting in Fig.~\ref{trajectory3}, where the desired relative positions between each agent and the target are specified to be $[-7, -7]\t, [7, -7]\t, [7, 7]\t, $ $[-7, 7]\t$ in advance. It is observed in Figs.~\ref{fixed_trajectory3} (a)-(d) that the rigid-formation fencing of agents from initial states (blue circles) fulfilling condition C1 is achieved with the same labels and relative positions (yellow circles). 
Comparing Figs.~\ref{trajectory3} (a)-(d) with Figs.~\ref{fixed_trajectory3} (a)-(d), there exist more oscillations in label-fixed fencing in Fig.~\ref{fixed_trajectory3}, which verifies the flexibility and high efficiency of the label-free fencing method in terms of shrinking the moving distance. 
Fig.~\ref{sudden_broken3} (a) shows the robustness of the present label-free fencing controller \eqref{control_i1}, where agent~$4$ breaks down at $t=20$s, and agents $1,2,3$ can still form a triangular formation to fence the moving target. By contrast, the fencing mission is failed by the label-fixed strategy in Fig.~\ref{sudden_broken3} (b). The robustness of the label-free design is thus verified when experiencing agent breakdown.

As for the states evolution during the label-free fencing simulation, we take Fig.~\ref{trajectory3}~(b) as an example. 
Fig.~\ref{distance3} describes the states evolution of simulation in Fig.~\ref{trajectory3}~(b), where $\lim_{t\rightarrow\infty}\bar{e}(t)=0$ achieves $P_{x_d}(x)=0$ implicitly in P1. The pairwise distances 
among the agents keep $\|x_i-x_k\|>2, i\neq k, i, k \in\mathcal V$, which verifies the inter-agent collision avoidance with $r=2$. 
To quantitatively show the improvement of the label-free approach, Fig.~\ref{fixed_distance3} illustrates the corresponding states evolution of Fig.~\ref{fixed_trajectory3}~(b) with the label-fixed strategy. More precisely, the fencing errors between the proposed label-free fencing controller~(9) (see Fig.~\ref{distance3} (a)) and the label-fixed strategy (see Fig.~\ref{fixed_distance3} (a)) both converge to zeros in less than $110$ seconds, which implies that the fencing efficiencies of these two approaches are almost the same. However, comparing the pairwise distances among the agents in Figs.~\ref{distance3}~(b) and \ref{fixed_distance3} (b), the highest amplitude oscillation of $\|x_i-x_j\|, i\neq k, i, k \in\cal V$ governed by the label-fixed strategy in Fig.~\ref{fixed_distance3} (b) is 60\%
 larger than such oscillation with the label-free approach in Fig.~\ref{distance3}~(b). Moreover, the convergence time of $\|x_i-x_j\|, i\neq k, i, k \in\cal V$ in Fig.~\ref{fixed_distance3}~(b) is 50 seconds longer than the corresponding time in Fig.~\ref{distance3}~(b). Both of them verify the superiority of the label-free fencing controller in terms of  the substantially shrinked moving distance and forming time to a rigid formation. 
Fig.~\ref{center3} exhibits that $\lim_{t\rightarrow\infty}x_i(t)-x_d(t)=d_i\neq0, \lim_{t\rightarrow\infty}v_i(t)-v_d(t)=0$ of the simulation in Fig.~\ref{trajectory3} (b), which implies the forming of the rigid formation fulfilling  Definition~\ref{rigid}. The feasibility of Theorem~\ref{controlaw1} is thus verified.

\section{Conclusion}
In this paper, we propose a label-free control scheme such that second-order MASs are capable of cooperatively fencing a moving 
target of variational velocity within a convex hull. Moreover, inter-agent collision avoidance and rigid formation are both
guaranteed without predeterminingly labeling any specified agents. The developed cooperative control protocol has been substantiated by numerical 
simulations. 
The future work may include extension of the proposed method for higher-dimensional systems and a more general target
such as 3D label-free fencing with general $S$ matrix.

\bibliographystyle{IEEEtran}
\bibliography{IEEEabrv,ref}

\end{document}